\documentstyle{article}
\renewcommand{\theequation}{\arabic{section}.\arabic{equation}}
\renewcommand{\baselinestretch}{1.5}

\def\Journal#1#2#3#4{{#1} {\bf #2}, #3 (#4)}

\def\NPB{{\em Nucl. Phys.} B}

\def\PRL{\em Phys. Rev. Lett.}
\def\PRB{{\em Phys. Rev.} B}

\def\JPC{{\em J. Phys.}C}

\newcommand{\no}{\nonumber\\}
\newcommand{\beg}{\begin{eqnarray}}
\newcommand{\en}{\end{eqnarray}}
\newcommand{\bef}{\begin{figure}}
\newcommand{\enf}{\end{figure}}

\newcommand{\udp}{\underline{dp}}
\newcommand{\udq}{\underline{dq}}
\newcommand{\udo}{\underline{d\omega}}
\newcommand{\ude}{\underline{d\epsilon}}
\newcommand{\ol}{\overline}

\newcommand{\olo}{\ol{\omega}}
\newcommand{\ole}{\ol{E}}

\newcommand{\olk}{\ol{\kappa}}

\newcommand{\FL}{F_{\Lambda}}

\newcommand{\sign}{\mbox{sgn}}
\newcommand{\ky}{\kappa,y}
\newcommand{\oky}{\olk,y}

\newcommand{\oM}{\olk^2+y^2}

\newcommand{\tilom}{\tilde{\omega}}
\newcommand{\tilep}{\tilde{\epsilon}}

\begin{document}
\renewcommand{\baselinestretch}{1.5}
\large
\vfill
\begin{flushright}
UT-Komaba 96-19
\end{flushright}

\begin{center}                      

{\LARGE{\bf Instability of the Non-Fermi-Liquid Fixed Point 
\\ in the Dissipative Gauge Theory of Fermions (I): \vskip 0.16in 
 Impurity Effects}}  
 \vskip 0.2in
{\Large Hiroshi Takano$^{\dagger}$\footnote
{e-mail address: takano@juen.ac.jp},}
{\Large  Masaru Onoda$^{\ddagger}$\footnote
{e-mail address: onoda@cms.phys.s.u-tokyo.ac.jp}, \\
Ikuo Ichinose$^{\ast}$\footnote{e-mail 
 address: ikuo@hep1.c.u-tokyo.ac.jp},
Tetsuo Matsui$^{\circ}$\footnote{e-mail
 address: matsui@phys.kindai.ac.jp}}  \\
$^{\dagger}$Joetsu University of Education,
Joetsu, Niigata, 943 Japan\\
$^{\ddagger}$Department of Physics, University of Tokyo,
Hongo, Tokyo, 113 Japan\\
$^{\ast}$Institute of Physics, University of Tokyo, 
Komaba, Tokyo, 153 Japan  \\
$^{\circ}$Department of Physics, Kinki University, 
Higashi-Osaka, 577 Japan
\end{center}
\begin{center}
\begin{bf}
Abstract
\end{bf}
\end{center}  
We study  a dissipative gauge theory of 
nonrelativistic fermions in  2+1  dimensions at zero temperature 
by the Wilsonian renormalization-group method. 
In this theory, we incorporate in the fermion propagator 
a new term of the form $ i \kappa \cdot\sign(\omega)$ 
(where $\kappa$ is a parameter and 
$\omega$ is the fermion frequency), which is usually induced by impurity effects.
In the previous papers, we studied this system for $\kappa = 0$, and 
showed that there exists a non-Fermi-liquid infrared fixed point.
In this paper, we address the question  whether this non-Fermi-liquid behavior
remains stable or not in the presence of impurity effects, i.e., the $\kappa$
term.
Our results show that the non-Fermi-liquid fixed point is unstable for 
$\kappa \neq 0$ and 
an effective gauge coupling constant tends to vanish at low energies.
However, in intermediate energy scales, the behavior of correlation functions
for $\kappa \neq 0$ is controlled
by the non-Fermi-liquid fixed point at $\kappa=0$, i.e., a crossover phenomenon
appears.
Physical implications of this phenomenon are discussed.

\vfill

\eject
%
%
\section{Introduction}
\setcounter{equation}{0}
%
In  condensed matter physics, low-dimensional electron systems 
are of current interest, which include the high-$T_c$ superconductivity,  
the fractional quantum Hall effect, and the recently discovered 
new spin ladder systems which exhibit superconductivity.
Theory  of  nonrelativistic fermions 
interacting with  a gauge field  plays  an important role as an effective
reference theory for these interesting systems.
A typical example of such gauge theory is found in
the work of Halperin, Lee, and Read\cite{halperin},
in which they studied a system of electrons in the half-filled Landau level.
This gauge theory is also regarded as an important constituent to compose a  
low-energy effective field theory for  the t-J model\cite{Lee}.

Recently, it has been recognized that the renormalization-group (RG)
method can be  very efficiently applicable to  a system of 
nonrelativistic fermions for investigating its low-energy 
behavior\cite{shanker}. 
Nayak and Wilczek\cite{Nayak} made a RG analysis of a kind of $\epsilon$-expansion 
for the above gauge theory 
in Ref.\cite{halperin} at zero temperature, and claimed existence of
a nontrivial infrared (IR) fixed point in some parameter region.
Following their works there  appeared some related works\cite{ichinose,sonota}.
This fixed point is very interesting because it implies a departure from the
conventional Fermi liquid theory; At that fixed point, 
the theory exhibits non-Fermi-liquid-like behavior similar to that of
a  marginal Fermi liquid for the high-$T_c$ superconductivity\cite{Varma} or 
of a Luttinger liquid in one dimension\cite{Hald}.

In this theory, a dissipative term is generated in the gauge-field  propagator
via    quantum fluctuations of fermions, and this term becomes important
  at low energies. Thus its effect to the above
nontrivial fixed point should be clarified. In the Nayak and Wilczek approach,
the relevance of this term to the above fixed point is not reflected well,
since the dissipative term has no divergence in their $\epsilon$-expansion,
hence not being treated as a relevant coupling. 
In the previous paper\cite{onoda}, we considered a two-dimensional model of nonrelativistic fermions interacting with 
a  gauge field having a dissipative term from the beginning, and 
carried out the Wilsonian RG (WRG) analysis, treating this dissipative term as a 
genuine running couping constant.  We found that there still exists a nontrivial fixed point. In this sense, the fixed point found by Nayak and Wilczek is stable
against the dissipative term.

Since this fixed point is so important, it is worth  studying its stability
from a more general point of view. In this paper, 
we study its stability against  ``impurities" in detail.      
Explicitly, we  incorporate an additional term in the fermion propagator of the previous dissipative gauge theory of Ref.\cite{onoda},
and study its  low-energy effects by the  WRG method.
The term added is 
$$
-i \kappa \cdot \sign (\omega),
$$
where $\kappa$ is a parameter, $\omega$ is the fermion frequency, and 
\beg
\sign(x)=\cases{1 & ($x>0$) \cr
               -1 & ($x<0$)  \cr }.
\en
Motivation for adding  the above term is twofold. 
First, this term is just generated via scattering of fermions by impurities.
Real physical systems necessarily  contain  impurities, and they often
play an important role; for example, in localization phenomenon 
in the quantum Hall effect.
It is  natural to extend the previous analysis in 
Ref.\cite{onoda} to a system containing impurities.
Usually, one treats  effects of random 
impurities by letting them to interact with fermions through a potential
 $v(x)$, so the interaction Hamiltonian is given by
\beg
H_{int} = \int dx \psi^{\dagger}(x) \sum_i v(x-x_i) \psi(x),
\en
where $\psi(x)$ is the fermion field operator and $x_i$ are coordinates of impurities.
Practical calculations at  one-loop level show  that the self energy $\Sigma_k(\omega)$ of fermions
has the following signature term
\beg
\Sigma_k(i\omega)=-i \kappa \cdot\sign(\omega),
\en
where  $\kappa (> 0)$ is related with the Fourier modes of 
$v(x)$. In the classical equation of motion of a fermion, the resistive
(friction) force is given by $\propto -\kappa \vec{v}$, where $\vec{v}$ 
is the velocity of  fermion.
In the random phase approximation (RPA) with respect to impurities,
the  fermion propagator is simply given as
\beg
G_0(k,\omega) =\frac{1}{i\omega+i\kappa \sign(\omega)
+\mu-E_k},
\label{propa0}
\en
where $E_k$ is the energy of fermion with momentum $k$, 
and $\mu$ is the chemical potential.

The constant $\kappa ( > 0 )$ has another physical significance. 
This term modifies the momentum distribution of the fermion density  
which exhibits a step-function behavior at zero temperature $T=0$ 
and $\kappa=0$.
Actually, from the propagator (\ref{propa0}), the density of fermions $N(E_k)$ 
at $E_k$ is obtained as 
\beg
 N(E_k) &=& \int_{-\infty}^{\infty} \frac{d\omega}{2 \pi}
\frac{1}{i\omega+i\kappa \sign (\omega)+\mu-E_k} +
\frac{1}{2}\no
&=&
\sign(\mu-E_k) \Bigl(\frac{1}{2}-\frac{1}{\pi}\arctan 
\frac{\kappa}{|\mu-E_k|}\Bigl)+\frac{1}{2}.
\label{N}
\en
The curve $N(E_k)$ is plotted  in Fig.1. From (\ref{N}) and Fig.1, 
we see that the $\kappa$ term removes
the singularity at $E_k = 0$. 

As is well known, this smooth momentum
distribution (\ref{N}) looks similar to that of a system of fermions at finite temperature ($T$).
The second motivation is related with this point. 
When we consider thermal  effects 
at finite $T$ in this model,   quantum corrections induce this 
signature term in the fermion self energy,  even if we start with 
$\kappa = 0${\bf }\cite{takano}.
Thus we must 
consider this signature term from the beginning  when we investigate the
system at $T > 0$ by using  the WRG analysis, treating $\kappa$ 
as a running coupling constant.
These motivations let us  start with the fermion action  
with the signature term as given by Eq.(\ref{ffermion}) in Sec.2.

As we explained above, in the case of $\kappa=0$ there is the nontrivial 
IR fixed point
at which the system exhibits non-Fermi-liquid-like behavior\cite{onoda,Nayak}.
We want to study the stability of this non-Fermi-liquid fixed point and 
the RG flow in the presence of the signature term.
Throughout the paper we consider the case of zero temperature ($T = 0$).
The other important case of finite $T$ will be discussed 
separately\cite{takano}.

This paper is organized as follows.
In Sec.2, we shall introduce the model.
In Sec.3, the RG prescription is explained.
We shall employ some specific momentum regularization, which is suitable for
the present gauge-fermion system.
In Sec.4, we shall calculate one-loop corrections of fermions and gauge bosons
by the momentum-shell integration.
In Sec.5, renormalization constants and RG equations are explicitly obtained.
Qualitative and numerical solutions of the RG equations are given.
It is shown that the non-Fermi-liquid fixed point at $\kappa=0$ is unstable
for nonzero $\kappa$.
However,  at certain intermediate energy scales, the behavior of the system is 
controlled by that fixed point. This means that the system exhibits a cross-over phenomenon.
In Sec.6, by solving the RG equation for the fermion propagator, we obtain 
its explicit form at intermediate energy scales and also in the low-energy limit,
confirming the above conclusion of the  cross-over.
Section 7 is devoted for concluding remarks.

\section{ The Model} 
\setcounter{equation}{0}
In this section we explain our model in detail, which  describes a 
2-dimensional  system of nonrelativistic spinless fermions $\psi(x,\tau)$ 
 interacting with  a dissipative gauge field $A_i(x,\tau)$ $(i=1,2)$,
where $0 < \tau <  \infty$ is the imaginary time.
 Relationships of this model with electrons in the half-filled Landau
 level and with strongly correlated electron systems for high-$T_c$ superconductivity
 are discussed in Refs.\cite{halperin,Lee}.
 
To study the low-energy excitations around the Fermi surface
and perform the WRG analysis for the gauge-fermion system,
it is useful to introduce a momentum cutoff as shown in Figs.2 and 3\cite{onoda}.
The region around the Fermi surface is divided into $N$ segments, and
the fermion field in each divided part is  labeled by $a$, $a=1,...,N$, 
as $\psi(a;p,\omega)$.
The fermion momentum $p$ is measured from 
a point on the Fermi surface $k_{F,a}$, the center of the $a$-th part.  
The momentum cutoff $\Lambda$ is introduced as
 $|p|=|k - k_{F,a}| < \Lambda$,
where $k$ is the original fermion momentum.
 
In the standard  path integral formalism, the partition
function of this model is given by 
\beg
Z &=& \int [d \bar{\psi}][d \psi][d A] \exp (-S_{\psi}-S_{A}-S_{int.3}).
\label{partition}
\en
In (\ref{partition}), $S_{\psi}$ is the free part of fermions 
given in the Fourier (momentum and energy) representation by 
\begin{eqnarray}
S_{\psi}&=&\int \udo \int \underline{dp}\;\sum_{a}
\bar{\psi}(a;p,\omega) (-i\omega -i \kappa \sign(\omega)
+v_F e_a \cdot p)
\psi(a;p,\omega) , 
\label{ffermion}
\en
where $\epsilon_a$ is the unit vector in the direction of $k_{F,a}$ and
\beg
\underline{dp} \equiv \frac{d^2p}{(2 \pi)^2}
,~~~\udo \equiv \frac{d\omega}{2 \pi}, ~~~v_F=k_F/m.
\en
The free part of  gauge field  $S_{A}$ is given by
\beg
S_{A}&=&\int\ude \int\underline{dq}   
 A(-q,-\epsilon)\Lambda^{b-1}v_B
\Bigl(q^{2-b}+\Lambda^{2-b}\lambda
\frac{|\epsilon|}{v_F q}
\Bigl)A(q,\epsilon) . 
\label{SA}
\en
The second term, with the coefficient $\lambda>0$, is the so-called dissipative term (Landau dumping factor)
that is induced by the  vacuum polarizations of fermions at one-loop level.
It is shown in the following discussion that the parameter $\lambda$ behaves 
nontrivially under RG transformation.
We have taken the Coulomb gauge $\partial_i A_i=0$, then the gauge field
$A_i$  has only one dynamical degrees of freedom, 
$A(q,\epsilon) \equiv i\epsilon_{ij} q_i A_j(q,\epsilon) /q$.
We have introduced a parameter $b$ which controls fluctuations of the gauge field.
This generalization was first
introduced in  Ref.\cite{halperin}.
The case $b=1$ corresponds to the Coulomb potential for electrons in the
half-filled Landau level, and the case $b = 0$ corresponds to the case of 
the t-J model.
The three-point interaction term of the action $S_{int,3}$ is given by 
\beg
S_{int,3}=igv_F\int \udo \ude \int\underline{dp}\;\underline{dq}\;  
\sum_{a}\frac{e_{a}\times q}{q} A(q,\epsilon)  
\bar{\psi}(a;p+q,\omega+\epsilon)\psi(a;p,\omega),
 \en
where $p\times q\equiv \epsilon_{ij}p_i q_j$.
In a nonrelativistic fermion-gauge field system, the action also contains the four-point
interaction term like $AA\bar{\psi} \psi$.  We shall neglect this term, 
  because this term contributes only to the
 renormalization
 of the chemical potential of the fermion and the effective mass
 of the gauge
field. Actually, in the field theoretical approach\cite{ichinose},
it is shown that the mass
of gauge field vanishes because the contribution from the 
four-point interaction cancels that from the three-point interaction, 
as guaranteed by the gauge invariance. 
In the following discussion,
we shall not discuss the contribution from these terms to the renormalization of the
chemical potential and the gauge field mass.

\section{ Wilsonian renormalization group program }

\setcounter{equation}{0}
In this section, we shall explain the WRG program for our gauge-fermion 
system introduced in Sec.2. 
In the previous paper\cite{onoda}, we studied the case $\kappa=0$ 
by the WRG {\bf }and showed that
there is a nontrivial IR fixed point for $b<1$.
Let us explain the procedures of  WRG program briefly once gain.
It consists of the following three steps (i)-(iii).

{\bf Step (i)} Integrate over the high-momentum modes (modes with the momentum
$\Lambda/e^t<p<\Lambda;t(>0)$ is a scaling parameter) of fermion and gauge fileds. 
Then the action of the low-momentum modes is obtained as follows;
\beg
\widetilde{S}_{\psi}
&=&
\int \udo \int^{\Lambda/e^t} 
\underline{dp}\no 
& &\times\sum_{a}
\bar{\psi}(a;p,\omega)R(t) 
\Bigl(-i\omega -i R_{\kappa}(t)\kappa \sign(\omega) 
+
R_{v_F}(t)v_F e_a \cdot p \Bigl)
\psi(a;p,\omega),
\no
\widetilde{S}_{A}
&=&
\int \ude \int^{\Lambda/e^t}
\underline{dq}\;   
 A(-q,-\epsilon)\Lambda^{b-1}v_B\Bigl(q^{2-b}+
\Lambda^{2-b}R_{\lambda}(t) \lambda
\frac{|\epsilon|}{v_F q}\Bigl)A(q,\epsilon),
 \no
\widetilde{S}_{int,3} 
&=&
\int \udo \ude
\int^{\Lambda/e^t}\underline{dp}\;\underline{dq} \no
& &\times \sum_{a}
iR_g(t) gv_F\frac{e_{a}\times q}{q} 
A(q,\epsilon)  
\bar{\psi}(a;p+q,\omega+\epsilon)\psi(a;p,\omega),
\en
where $R(t)$, etc. are $t$-dependent renormalization constants. 
We have neglected terms that are higher -orders in momentum or fields as  
irrelevant terms.
This point is discussed somewhat in detail in Ref.\cite{onodadh}.

{\bf Step (ii)} The original action $S$ and the new action 
$\tilde{S} =  \tilde{S}_{\psi} 
  +   \tilde{S}_A + \tilde{S}_{int,3}$ are defined
in the two different kinematical regions, and then we rescale momenta
in order to  return the new momentum cutoff $\Lambda/e^t$ to 
the original value $\Lambda$.
Similarly, we require so that certain quadratic terms 
in the effective action have the same coefficients with the original ones by rescaling fields. At the same time we scale energy, momentum and the fields.
In nonrelativistic theories like the present model, there is an ambiguity in assigning scaling
dimensions of energy and fields, because there is no principle for that. 
This is in sharp contrast with   
relativistic systems where  it is natural to assign  same scalings for
 energy and momentum. 
Therefore  we introduce a general scaling law which contains a 
free parameter $\xi>0 $ as follows,
\beg
\omega& \to &\tilde{\omega}=e^{\xi t}\omega,\no
p &\to& \tilde{p}=e^t p,\  (\ |p| \leq \Lambda /e^t, \ |\tilde{p}|\  \leq 
\Lambda )\no
\psi(a;p,\omega) &\to& \tilde{\psi}(a;\tilde{p},\tilde{\omega})
=e^{-(\xi + 1)t}\sqrt{R(t)}\psi(a;p,\omega),\no
A(p,\omega) & \to & \tilde{A}(\tilde{p},\tilde{\omega})
= e^{-(2+\frac{\xi-b}{2})t} A(p,\omega).
\label{scale}
\en  
{\bf Step (iii)}
The effective action $\tilde{S}$ is expressed in terms of the above variables; 
\begin{eqnarray}
\widetilde{S}_{\psi}
&=&
\int \underline{d\tilde{\omega}} \int^{\Lambda} 
\underline{d\tilde{p}}\sum_{a}
\bar{\tilde{\psi}}(a;\tilde{p},\tilde{\omega})(-i\tilde{\omega} -i \kappa(t) \sign(\tilde{\omega})
+v_F(t) e_a \cdot \tilde{p})
\tilde{\psi}(a;\tilde{p},\tilde{\omega}) ,\no
\widetilde{S}_{A}
&=&
\int \underline{d\tilde{\epsilon}} \int^{\Lambda}
\underline{d\tilde{q}}\;   
 \tilde{A}(-\tilde{q},-\tilde{\epsilon})\Lambda^{b-1}v_B\Bigl(\tilde{q}^{2-b}
+\Lambda^{2-b}\lambda(t)\frac{|\tilde{\epsilon}|}{v_F(t) 
\tilde{q}}\Bigl)\tilde{A}(\tilde{q},\tilde{\epsilon}),
\no
\widetilde{S}_{int,3}
 &=&\!\!\!
\int \underline{d\tilde{\omega}} \underline{d\tilde{\epsilon}} 
\int^{\Lambda}\underline{d\tilde{p}}\;\underline{d\tilde{q}}\;  
\sum_{a}ig(t)v_F(t)\frac{e_{a}\times \tilde{q}}{\tilde{q}} 
\tilde{A}(\tilde{q},\tilde{\epsilon})  
\bar{\tilde{\psi}}(a;\tilde{p}+\tilde{q},\tilde{\omega}+\tilde{\epsilon})
\tilde{\psi}(a;\tilde{p},\tilde{\omega}).
\en
The effective parameters (coupling constants) which flow as 
the scaling parameter
$t$ varies have been defined as follows,
\beg
g(t) &=& e^{(1-\frac{b+\xi}{2})t}\frac{R_g(t)}{R(t)R_{v_F}(t)}g,\no
v_F(t) &=& e^{(\xi-1)t}R_{v_F}(t)v_F,\no
\lambda(t) &=& e^{(2-b)t}R_{\lambda}(t)\lambda,\no
\kappa(t) &=& e^{\xi t}R_{\kappa}(t) \kappa.
\label{sl}
\en 
The simplest assignment of scaling dimensions is $\xi=1$.
One-loop calculation shows that RG recursion  equations 
of the effective parameters depend on the values of $\xi$.
But  this $\xi$-dependence must be superficial.
Actually, as we show in later discussion, physical quantities are independent of $\xi$.

\section{ One-loop quantum corrections and renormalization constants} 
\setcounter{equation}{0}

In this section, we shall perform the 
momentum-shell integration for an infinitesimal 
RG transformation with $\exp(dt) \simeq 1+dt$ at one-loop level
and calculate the renormalization
constants $R,R_{v_F},R_g,R_{\lambda}$ and $R_{\kappa}$.
With this result of  infinitesimal RG transformation, we  derive 
the RG equations of our running coupling constants,
$g(t),v_F(t),\lambda(t)$ and $\kappa(t)$. We consider the fermion part
in the action in Sec.4.1, the gauge-boson part in Sec.4.2, and 
the vertex part in Sec.4.3.

%
\subsection{Renormalization of the fermion part}
%

The bare action is renormalized by the one-loop quantum correction 
which is obtained by  integrating over the high-momentum modes of fermion 
field and  gauge field  in the momentum region $\Lambda(1-dt)
<p<\Lambda$.
The effective action is obtained as follows
\beg
\tilde{S}_{\psi}=\sum_{a}\int \udo \int_0^{\Lambda(1-dt)}\!\!\!
\udp \bar{\psi}(a;p,\omega)\Bigl[-i\omega-i \kappa \sign(\omega)
+v_F e_{a}\cdot p
 +\Sigma(a;p,\omega)\Bigl] \psi(a;p,\omega),
\en
where $\Sigma(a;p,\omega)$ is the fermion self-energy coming from
the high-momentum modes. It is given by
\beg
\Sigma(a;p,\omega) 
 =g^2 v_F^2\int \ude \int_{\Lambda(1-dt)}^{\Lambda}\udq
\left(\frac{e_{a}  \times q}{q}\right)^2 D_0(q,\epsilon)
G_0(a;p+q,\omega+\epsilon), 
\label{sigdef}
\en
where $D_0(q,\epsilon)$ is propagator of the gauge field obtained from
(\ref{SA}), 
\beg
D_0(q,\epsilon) &=& \frac{1}{\gamma(q)(|\epsilon|+F(q))}, \no
\gamma(q) &=& \lambda \frac{v_B \Lambda}{v_F q},\no 
F(q)&=&\gamma^{-1}(q)\Lambda^{b-1}v_Bq^{2-b}
=\frac{v_Fq}{\lambda}(\frac{q}{\Lambda})^{2-b}.
\en
Similarly,
$G_0(a;p,\omega)$ is  the fermion propagator,   
  obtained  from $S_{\psi}$ of (\ref{ffermion})   as
\beg
G_0(a;p,\omega)&=&\frac{1}{ i\omega +i\kappa \sign(\omega)
- E_p}, \no
 E_p &\equiv & v_F e_{a} \cdot p,
\label{fprop}
\en
with $\kappa >0$.

The  momentum-shell integration over the
 radial direction in Eq.(\ref{sigdef}) can be easily performed
to obtain the self-energy  as
\beg
\Sigma(a;p,\omega) 
 =-i \alpha F_{\Lambda} dt\int^{\lambda}_{-\lambda}
dy |\sin\phi (y)| I(\olo,\ole,\olk),
\label{selfenergy}
\en
where
\beg
I(\olo,\ole,\olk) \equiv \int^{\infty}_{-\infty}
d\bar{\epsilon} \frac{1}{\bar{\epsilon} +\olo +\olk \sign(\bar{\epsilon} +
\olo)
+i\ole} \cdot \frac{1}{|\bar{\epsilon}|+1}.
\label{I}
\en
In the above we have introduced  
\beg
\alpha &\equiv& \frac{g^2 v_F}{2 \pi^2 v_B},\no
F_{\Lambda} &\equiv& F(\Lambda)=\frac{v_F \Lambda}{\lambda}, \no
\overline{E} &\equiv & y + \frac{v_F e_a \cdot  q}{F_{\Lambda}}~,~~ y \equiv  
\frac{v_F \Lambda \cos\phi}{F_{\Lambda}}, \no 
\overline{\epsilon} &\equiv& \frac{\epsilon}{F_{\Lambda}}~,~~
\olo \equiv \frac{\omega}{F_{\Lambda}}~,~~
\olk \equiv \frac{\kappa}{F_{\Lambda}}.
\label{f1}
\en
The $\phi$ is the angle between the two vectors $p$ and $e_a$.
In the above momentum integration, we changed the integration variable  from
$\phi$ to $y$.

We divide the integrand $I$ (\ref{I}) into six terms as there are two signature 
terms $\sign(\overline{\epsilon} +\overline{\omega})$ and $|\overline{\epsilon}|$.  
\beg
I(\overline{\omega},\overline{E},\overline{\kappa})= \Theta(\overline{\omega}) \sum_{i=1}^{3}
I_i(\overline{\omega},\overline{E},\overline{\kappa}) - \Theta(-\overline{\omega}) \sum_{i=1}^{3}
I_i(-\overline{\omega},-\overline{E},\overline{\kappa}),
\label{idef1}
\en
where
\beg
I_1(\overline{\omega},\overline{E},\overline{\kappa}) &\equiv& \int_{-\infty}^{-\overline{\omega}}
d\overline{\epsilon} \frac{1}{\overline{\epsilon} +\overline{\omega} -\overline{\kappa} +i\overline{E}}
\cdot \frac{1}{-\overline{\epsilon}+1},\no
I_2(\overline{\omega},\overline{E},\overline{\kappa}) &\equiv& \int_{-\overline{\omega}}^{0}
d\overline{\epsilon} \frac{1}{\overline{\epsilon} +\overline{\omega} +\overline{\kappa} +i\overline{E}}
\cdot \frac{1}{-\overline{\epsilon}+1},\no
I_3(\overline{\omega},\overline{E},\overline{\kappa}) &\equiv& \int_{0}^{\infty}
d\overline{\epsilon} \frac{1}{\overline{\epsilon} +\overline{\omega} +\overline{\kappa} +i\overline{E}}
\cdot \frac{1}{\overline{\epsilon}+1},
\label{Is}
\en
and $\Theta(x)$ is the step function,
\beg
\Theta(x) = \cases{ 1 & ($x>0$) \cr
                    0 & ($x<0$) \cr} ~~.
\en
The integrations in (\ref{Is}) can be evaluated as
\beg
I_1(\overline{\omega},\overline{E},\overline{\kappa}) &=& -\frac{1}{z_2+1}
\Bigl[\frac{1}{2}\ln\frac{(\overline{\omega}+1)^2}
{\overline{\kappa}^2+\overline{E}^2}+ i\sign(\overline{E})(\frac{\pi}{2}-
\arctan\frac{\overline{\kappa}}{|\overline{E}|}) \Bigl],\no
I_2(\overline{\omega},\overline{E},\overline{\kappa}) &=& \frac{1}{z_1+1}
\Bigl[\frac{1}{2}(\ln\frac{(\overline{\omega}+\overline{\kappa})^2
+\overline{E}^2}{\overline{\kappa}^2+\overline{E}^2}
+\ln|\overline{\omega}+1|^2)
\no& &- i\sign(\overline{E})(\arctan\frac{\overline{\omega}
+\overline{\kappa}}{|\overline{E}|}-
\arctan\frac{\overline{\kappa}}{|\overline{E}|}) \Bigl],\no
I_3(\overline{\omega},\overline{E},\overline{\kappa}) &=& \frac{1}{z_1-1}
\Bigl[\frac{1}{2}(\ln((\overline{\omega}+\overline{\kappa})^2
+\overline{E}^2)+ i\sign(\overline{E})(\frac{\pi}{2}
-\arctan\frac{\overline{\omega}+\overline{\kappa}}{|\overline{E}|}) 
\Bigl],
\en
where
\beg
z_1 \equiv \overline{\omega} + i\overline{E} +\overline{\kappa},~~~
z_2 \equiv \overline{\omega} + i\overline{E} -\overline{\kappa}.
\label{idef2}
\en

We are interested in the renormalization constants, so we expand the 
self-energy $\Sigma(a;p,\omega)$ in powers of  $i \omega$ and $E_p \equiv
v_Fe_a\cdot p$ and keep only the linear terms;
\beg
\Sigma(a;p,\omega) = \Sigma_0 +
\Sigma_{\omega} \cdot i \omega + \Sigma_{p} \cdot E_p.
\en
Detailed calculations are given in Appendix A, and 
the expansion coefficients are obtained as follows; 
\beg
\Sigma_0 &=& 0, \no
\Sigma_{\omega} &=&-\alpha dt K_1(\kappa_v,\lambda),\no
\Sigma_p &=&-\alpha dt \Bigl(K_2(\kappa_v,\lambda)
-K_1(\kappa_v,\lambda)\Bigl),
\label{ecoeff}
\en
where we introduced the parameter $\kappa_v$, a rescaled $\kappa$, and two functions $K_{1,2}(\kappa_v, \lambda )$; 
\beg
\kappa_v &\equiv & \kappa/(v_F \Lambda), \no
K_1(\kappa_v,\lambda) &\equiv& 
\frac{1}{\pi} \int^1_0 dx \sqrt{1-x^2} \frac{1}{\lambda(\nu^2+x^2)}
\no
&\times&\Bigl[ 
2\lambda \nu - \frac{\nu^2-x^2}{\nu^2+x^2}
\ln\lambda^2(\kappa_v^2+x^2)-
\frac{4\lambda x \nu}{\nu^2+x^2}(\frac{\pi}{2}-
\arctan\frac{\kappa_v}{|x|}) \Bigl] ,\no 
K_2(\kappa_v,\lambda) 
&\equiv &
\frac{2\kappa_v}{\pi} \int^1_0 dx 
\frac{\sqrt{1-x^2}}{\kappa_v^2+x^2},  \no
\nu & \equiv & \kappa_v-{1\over \lambda}.
\label{K1K2}
\en
Below we mainly use $\kappa_v$ instead of $\kappa$.
We find that no terms proportional to $\sign(\omega)$
appear in the above expansion, and so the renormalization constant  of $\kappa$, $R_{\kappa}(t)$, equals to unity. 
Therefore the renormalization of  $\kappa$ comes only from the wave function renormalization $R(t)$.
From (\ref{ecoeff}), we obtain the renormalization constants 
for infinitesimal RG transformation as follows
\beg
R_{\kappa}(dt) &=& 1,\no
R(dt) &=& 1-\Sigma_{\omega}\no
& =&1 + \alpha dt K_1(\kappa_v,\lambda), \no
R_{v_F}(dt) &=&1 + \Sigma_{\omega}+\Sigma_{p} \no
&=&1 - \alpha dt K_2(\kappa_v,\lambda).
\label{rec}
\en
In Eq.(\ref{K1K2}) the dominant contributions to the integrals for $K_1$ and $K_2$ come from the region
$x \simeq 0$, so we can approximate as 
$\sqrt{1-x^2} \simeq 1$ in the integrand.
Then the $x$ integrations can be performed readily as
\beg
K_1(\kappa_v,\lambda) &\simeq&
\frac{2}{\pi}\arctan\frac{1}{\kappa_v}
-\frac{1}{\pi \lambda(1+\nu^2)}
\Bigl(\ln\lambda^2+\ln(\kappa_v^2+1)-
2\nu \arctan\frac{1}{\kappa_v}\Bigl) ,\no
K_2(\kappa_v) &\simeq& \frac{2}{\pi}\arctan\frac{1}{\kappa_v}.
\label{K1}
\en

%
\subsection{Renormalization of the gauge-field part}
%
By integrating
the high-momentum modes, we obtain the one-loop correction 
to the action $S_A$,  so the renormalized action  $\tilde{S}_{A}$ becomes 
\beg
\tilde{S}_{A}=\int \ude \int^{\Lambda(1-dt)}
\udq A(-q,-\epsilon)\Bigl[ \Lambda^{1-b}v_B
(q^{2-b}+\Lambda^{2-b}\lambda\frac{|\epsilon|}{v_F q})
+\Pi(q,\epsilon) \Bigl] A(q,\epsilon),
\label{sa}
\en
where the vacuum polarization $\Pi(q,\epsilon)$ reads explicitly as
\beg
\Pi(q,\epsilon)=g^2v_F^2 \sum_a 
\Bigl(\frac{e_a \times q}{q}\Bigl)^2
\int \udo \int^{\Lambda}_{\Lambda(1-dt)}
\udp \; G_0(a;p+q,\omega+\epsilon) G_0(a;p,\omega).
\en
After doing the integration in the  radial direction of $q$,
$\Pi(q,\epsilon)$ is expressed as 
\beg
\Pi(q,\epsilon) = \frac{g^2 v_F \Lambda dt}{2\pi^2}
 \sum_a 
\Bigl(\frac{e_a \times q}{q}\Bigl)^2
\int^{v_F \Lambda}_{-v_F \Lambda} dy
\frac{1}{|\sin \phi(y)|}
P(\epsilon,E_q,y,\kappa),
\label{selfenergy2}
\en
where
\beg
P(\epsilon,E_q,y,\kappa) &\equiv&
   \Theta(\epsilon) \sum_{i=1}^3 P_i(\epsilon,E_q,y,\kappa)
+\Theta(-\epsilon) \sum_{i=1}^3P_i(-\epsilon,-E_q,-y,\kappa)
 ,
\en
with
\beg
P_1(\epsilon,E_q,y,\kappa)
&\equiv&
\int_{-\infty}^{-\epsilon} \udo
\frac{1}{i(\omega+\epsilon)-i\kappa -y-E_q}
\frac{1}{i\omega - i\kappa -E_q}, \no
P_2(\epsilon,E_q,y,\kappa)
&\equiv&
\int_{-\epsilon}^{0} \udo
\frac{1}{i(\omega+\epsilon)+i\kappa -y-E_q}
\frac{1}{i\omega - i\kappa -E_q}, \no
P_3(\epsilon,E_q,y,\kappa)
&\equiv&
\int_{0}^{\infty} \udo
\frac{1}{i(\omega+\epsilon)+i\kappa -y-E_q}
\frac{1}{i\omega + i\kappa -E_q}.
\en
We have changed the integration variable  from $\phi$ to $y$ as before.
The above integrations can be evaluated as follows;
\beg
P_1(\epsilon,E_q,y,\kappa)&=&
\frac{i}{2 \pi(i\epsilon-E_q)}
\Bigl[ \frac{1}{2}\ln
\frac{\kappa^2+E^2}{(\epsilon+\kappa)^2+E^2}
-i \sign(E)(\frac{\pi}{2}-\arctan\frac{\kappa}{|E|})
\no& &+
i \sign(y)(\frac{\pi}{2}-\arctan\frac{\epsilon+\kappa}{|y|})
\Bigl],
\no
P_2(\epsilon,E_q,y,\kappa)&=&
\frac{i}{2 \pi(i(\epsilon+2\kappa)-E_q)}
\Bigl[ \frac{1}{2}(\ln
\frac{(\epsilon+\kappa)^2+E^2}{\kappa^2+E^2}
-
\ln
\frac{\kappa^2+E_p^2}{(\epsilon+\kappa)^2+y^2})
\no& &-
i \sign(E)(\arctan\frac{\epsilon+\kappa}{|E|}
-\arctan\frac{\kappa}{|E|})\no
& &+i \sign(y)(\arctan\frac{\kappa}{|y|}
-\arctan\frac{\epsilon+\kappa}{|y|})
\Bigl],
\no
P_3(\epsilon,E_q,y,\kappa)&=&
\frac{i}{2 \pi(i\epsilon-E_q)}
\Bigl[ \frac{1}{2}\ln
\frac{(\epsilon+\kappa)^2+E^2}{\kappa^2+y^2}
\no& &-
i \sign(E)(\frac{\pi}{2}-\arctan\frac{\epsilon+\kappa}{|E|})
+i \sign(y)(\frac{\pi}{2}-\arctan\frac{\kappa}{|y|})
\Bigl],
\en
where $E = E_q + y $.
We are interested in the low-energy modes, so we expand these results
in powers of $q$ and $E_q$. 
Detailed calculations are given in Appendix  B. 
Explicitly, for $ |\epsilon| /(v_F q) << 1$ \cite{ichinose,onoda}
 we obtain 
\beg
\Pi(q,\epsilon) &=& \frac{g^2 v_F \Lambda dt}{2 \pi^2}
\sum_a \frac{K_a E_q}{2\pi(i\epsilon-E_q)}
\int_0^1dx \frac{4}{\sqrt{1-x^2}}
\frac{\kappa_v}{\kappa_v^2+x^2}\no
&\simeq&
v_B \Lambda  \alpha dt
\frac{\pi k_F}{\Lambda} \frac{|\epsilon|}{v_F q}
\frac{2}{\pi}\arctan\frac{1}{\kappa_v},
\label{vacpol}
\en
where
\beg
K_a &\equiv& \Bigl(\frac{e_a \times q}{q} \Bigl)^2
\en
and we set
 $\sqrt{1-x^2} \simeq 1$ in the $x$-integration above as before.
This result has a form of the dissipative ($\lambda$) term. 
We neglected a constant term, which is a mass renormalization of  gauge field, 
because this term is canceled by the contribution from 
the four-point interaction as explained in the introduction.

By substituting these results into Eq.(\ref{sa}), we
get the infinitesimal renormalization constant for
the coefficient of the dissipative term,
\beg
R_{\lambda}(dt) \lambda  = \lambda + \alpha dt K_3(\kappa_v),
\label{rl}
\en
where
\beg
K_3(\kappa_v) \equiv \frac{2k_F}{\Lambda}\arctan
\frac{1}{\kappa_v}.
\label{K3}
\en

%
\subsection{Vertex correction }
%

The renormalization of the vertex part is similarly obtained by the momentum-shell integration,
and the action $\tilde{S}_{int.3}$ is expressed as 
\beg
\tilde{S}_{int.3}= igv_F \int \udo \ude
\int^{\Lambda(1-dt)}\udp \udq &\sum_a& 
\frac{e_a \times q}{q}
\Bigl(  1+ \Gamma_3(a;p,\omega,p+q,\omega+\epsilon) \Bigl) \no
&\times& 
A(q,\epsilon) \bar{\psi}(a;p+q,\omega+\epsilon) \psi(a;p,\omega),
\en
where the vertex correction $\Gamma_3$ is given by
\beg
\Gamma_3(a;p,\omega,p',\omega') 
&= &
 \int^{\Lambda}_{\Lambda(1-dt)} \udq'
\frac{g^2 v_F^2}{\gamma(q')}
 \Bigl(\frac{e_a \times q'}{q'} \Bigl)^2 
\no &\times&
U(E_p+E_{q'},\omega;E_{p'}+E_{q'},\omega';F(q')),
\en
and
\beg
&& U(E,\omega;E',\omega';F)\no
&\equiv&
\int \ude'
G_0(a;E,\omega+\epsilon') 
G_0(a;E',\omega'+\epsilon')
\frac{1}{|\epsilon'|+F} \no
&=&
\int \ude'\Bigl(
G_0(a;E,\omega+\epsilon') -G_0(a;E',\omega'+\epsilon')
\Bigl) \frac{1}{|\epsilon'|+F}
\no
&\times&
\frac{1}{i(\omega'-\omega)+i\kappa (\sign(\omega'+\epsilon')
-\sign(\omega+\epsilon') )
-(E'-E)}.
\en
The corrected three-point coupling has a momentum dependence.
However, we take the ``on-shell" limit since
we are not interested in irrelevant higher-order terms
of the Taylor-expansion in powers of momenta and energies.
The on-shell limit of $\Gamma_3$ should be taken as \cite{ichinose}
\beg
\lim_{p,p' \to 0} \lim_{\omega,\omega' \to 0}
\Gamma_3(a;p,\omega,p',\omega')
&=&\lim_{p,p' \to 0} \frac{1}{E_p-E_{p'}}
(\Sigma(0,E_p)-\Sigma(0,E_p')+ O(E_p^2,E_{p'}^2)) \no
&=& \Sigma_p,
\en
where $\Sigma_p$ is given by Eq.(\ref{ecoeff}).
Therefore we obtain the renormalization constant $R_g(dt)$,
\beg
R_g(dt) &=& 1 + \Sigma_p \no
&=& R(dt) R_{v_F}(dt).
\label{rg}
\en
This relationship between $R_g = R  R_{v_F} $ is nothing but 
the Ward-Takahashi identity which is discussed in Ref.\cite{ichinose}.

%
%
\section{ WRG equations and the RG flows }
%
%
\setcounter{equation}{0}

In the previous section, we obtained infinitesimal forms of renormalization constants,
 Eqs.(\ref{rec}), (\ref{rl}) and (\ref{rg}).
These results and the  scaling laws (\ref{sl}) give the 
differential equations for the running parameters, 
$g(t), v_F(t) ,\lambda(t)$ and $\kappa_v(t)$ 
as follows,
\beg
\frac{dg(t)}{dt} &=& \frac{2-b -\xi}{2}g(t) , \no
\frac{dv_F(t)}{dt} &=& \Bigl(\xi-1-\frac{g(t)^2v_F(t)}{2\pi^2 v_B}
 K_2(t) \Bigl)v_F(t), \no
\frac{d\lambda(t)}{dt} &=&
 \Bigl( 2-b- \frac{g(t)^2v_F(t)}{2\pi^2 v_B}
K_2(t) \Bigl)\lambda(t) + 
\frac{g(t)^2v_F(t)}{2\pi^2 v_B}K_3(t),\no
\frac{d\kappa_v(t)}{dt} &=&
\Bigl(1-\frac{g(t)^2v_F(t)}{2\pi^2 v_B}(K_1(t)-K_2(t) )
\Bigl)\kappa_v(t),
\label{rge}
\en
where $K_i(t)$ ($i=1,2$ and $3$) means $K_i(\kappa(t),\lambda(t))$ 
given by (\ref{K1}) and (\ref{K3}).

In these equations, in place of the gauge coupling $g$ itself,
one can regard the following combination $\gamma$
as an effective expansion parameter in the perturbative calculation of the present theory,
\beg
\gamma(t) &\equiv& \frac{g(t)^2v_F(t)}{2\pi^2 v_B}K_2(t).
\label{a}
\en
In Ref.\cite{onoda} we have introduced a similar combination ( we called it
 $\alpha  = g(t)^2v_F(t) (2\pi^2 v_B)^{-1}$ there),  where
$\kappa=0$ and so $K_2(t)=1$.
This $\gamma$ is a generalization of it  in the case of  $\kappa > 0$.
The reason why the extra factor $K_2(t)$ appears is physically obvious from
(\ref{N}) and Fig.1.
That is, the density of states around the Fermi surface  decreases 
by the existence of the
$\kappa$-term as $N(E_k)\sim 1/2 \pm  (\arctan \;
{ \kappa_v^{-1}})/\pi  \sim  (1 \pm   K_2 )/2$, 
and so the interaction between  the  gauge field and fermions is weakened
by the factor $K_2(t)$.

Then we obtain  the following coupled differential
equations,
\beg
\frac{d\gamma(t)}{dt}&=&\Bigl[1-b-L_1(t)-
(1-L_1(t)L_2(t))\gamma(t) \Bigl] \gamma(t),
\label{gamma(t)} \\
\frac{dv_F(t)}{dt}&=& \Bigl( \xi-1-\gamma(t)\Bigl)v_F(t), \\
\frac{d\kappa_v(t)}{dt}&=&
\Bigl( 1-L_2(t)\gamma(t) \Bigl) \kappa_v(t),
\label{k(t)} \\
\frac{d\lambda(t)}{dt}&=&
\Bigl(2-b-\gamma(t)\Bigl)\lambda(t)+\frac{\pi k_F}{\Lambda}\gamma(t),
\label{la(t)}
\en
where
$L_1(t)$ and $L_2(t)$ mean the following $L_1(\kappa_v(t),\lambda(t))$ and
$L_2(\kappa(t),\lambda(t))$, respectively;
\beg
L_1(\kappa_v,\lambda)
&\equiv&
\frac{\kappa_v}{(\kappa_v^2+1) \arctan
\kappa_v^{-1}},\no
L_2(\kappa_v,\lambda)
&\equiv&
\frac{-1}{2 \pi \lambda(1+\nu^2)\arctan\kappa_v^{-1}}
\Bigl(\ln\lambda^2+\ln(1+\kappa_v^2)-2\nu\arctan\kappa_v^{-1}
\Bigl).
\en
For latter discussions on the IR behavior of the
parameters, it is somewhat useful to refer to the system with $\kappa_v = 0$
and $K_2 = 1$. Therefore we rewrite below these differential equations 
in terms of the combination,   
\beg
\alpha(t) &\equiv & \frac{\gamma(t)}{K_2(t)} = \frac{g(t)^2v_F(t)}{2\pi^2 v_B}, 
\label{alpha}
\en
as follows;
\beg
\frac{d\alpha(t)}{dt} &=& \Bigl(1-b-K_2(t)\alpha(t)
\Bigl)\alpha(t),
\label{alp(t)}\\
\frac{d\kappa_v(t)}{dt} &=& \Bigl( 1-(K_1(t)-K_2(t))
\alpha(t) \Bigl) \kappa_v(t),
\label{k(t)2}\\
\frac{d\lambda(t)}{dt} &=&
\Bigl(2-b-K_2(t)\alpha(t)\Bigl)\lambda(t) + K_3(t)\alpha(t).
\label{l(t)2} 
\en

The   equations (\ref{gamma(t)}), (\ref{k(t)}) and (\ref{la(t)})  
are a set of closed 
differential equations with respect to  the coupling constants,
$\gamma(t), \kappa_v(t)$ and $\lambda(t)$,  and they
do not depend on the scaling parameter $\xi$.
The behavior of $v_F(t)$ is calculated from these quantities.
Then, in the following discussions, we shall consider RG flows of only 
these  couplings.

As stated, we want to know whether the present
 theory including the $\kappa$-term
has a nontrivial fixed point or not. In the case 
$\kappa_v=0$, this theory has the nontrivial IR fixed point in the
parameter region $b<1$ such as $\gamma(\infty)=\gamma^*_0=1-b>0$.
The existence of this fixed point can be seen in Eq.(\ref{alp(t)}) 
by simply setting $K_2(\kappa_v=0)=1$, where $k_v =0$ is certainly
a solution.
We shall discuss how the existence of the
 $\kappa$-term influences the stability of this fixed point.

A formal solution to the differential equations
 (\ref{gamma(t)}), (\ref{k(t)}) and (\ref{la(t)})
is obtained as 
\beg
\gamma(t) &=& \frac{e^{(1-b)t}K_2(t)}
{\int_0^t ds e^{(1-b)s}K_2(s) + \alpha_0^{-1}},\no
\kappa_v(t) &=& \kappa_{v0} 
\exp(t-\int_0^t ds L_2(s) \gamma(s)),\no
\lambda(t) &=& \lambda_0 e^{(2-b)t-\int_0^t ds
\gamma(s)}+\frac{\pi k_F}{\Lambda}.
\label{formalsol}
\en
One  can determine the qualitative IR behavior of the parameters 
by the following arguments;   \\
(i) First, let us assume $\kappa_v$ has a finite constant value in the IR
limit, i.e., $\lim_{t\to\infty}\kappa_v(t) = \kappa_v^*>0$.
Then, from Eq.(\ref{alp(t)}), $\alpha(t)$ has the following fixed points $\alpha^*$,
\beg
 \alpha^*=
\cases{    \frac{1-b}{K_2(\kappa_v^*)} & $(b<1)$  \cr
                  0 & $(1 \le b \le 2)$ \cr } ~~.
\en
Then $\lambda(t)$ becomes large according to (\ref{l(t)2}) as 
\beg
\dot{\lambda}&\simeq&(2-b-K_2\alpha^*)\lambda + K_3 \alpha^*\no
&=&
\cases{\lambda+K_3 \alpha^* & $>0$  $(b<1)$ \cr
     (2-b)\lambda + K_3 \alpha^* & $>0$ $(1 \le b \le 2)$, \cr }
\en
where $\dot{\lambda} \equiv  d\lambda(t)/dt $.
Since $\lambda$ becomes large, the factor $K_1-K_2$ behaves like,
\beg
K_1-K_2 \simeq -\frac{\ln\lambda^2}{\pi \lambda} \to 0.
\en
This fact and Eq.(\ref{k(t)2}) indicate that $\kappa_v$ increases
exponentially. 
This is an obvious contradiction to  the starting assumption.
This implies that $\kappa_v$ cannot have a finite value in the IR limit.

(ii) From (i) we conclude that  the parameter $\kappa_v$ tends to large in
the IR limit. Thus $L_1$ goes to unity and $L_2$ is given as follows,
\beg
L_2 &\simeq& - \frac{1}{\kappa_v^{-2} x^2+(x-1)^2}
(x\ln x+1-x),
\en
where $x\equiv\kappa_v \lambda$.
We note that the  value of $L_2$ is restricted to  
  $-1 \le L_2 \le 0$.
The coupled equations are approximated as follows,
\beg
\dot{\gamma} &\simeq& \Bigl( -b+(L_2-1)\gamma \Bigl)\gamma,\no
\dot{\kappa_v} &\simeq& (1-L_2 \gamma)\kappa_v,\no
\dot{\lambda} &\simeq& (2-b-\gamma)\lambda + 
\frac{\pi k_F}{\Lambda}\gamma,
\en
where we have omitted the argument $t$ in  $\gamma(t)$ etc.
In (ii-1,2,3) below we argue that $\gamma(t) \rightarrow 0$ in the IR limit.

(ii-1)  Let us assume that  $\gamma(t)$ has a nontrivial fixed point
$\gamma^* = \frac{b}{L_2-1} > 0$. Since $\gamma(t)$ should be positive, 
$L_2$ should become  larger than unity. 
However  we found that $L_2 <0$ in (ii) above, and so this cannot happen. 

(ii-2) Let us assume that $\gamma(t)$ tends to  infinity.
Then, $\lambda(t)$ satisfies the
equation $\dot{\lambda} \simeq -\gamma\lambda +
  (\pi k_F/\Lambda ) \gamma$,
and we obtain $\lambda(t)$ as
\beg
\lambda(t) =
\exp(-\int_0^{\infty}ds \gamma(s))+\frac{\pi k_F}{\Lambda}.
\en
From the behavior of $\kappa_v(t)$ and $\lambda(t)$ at large $t$,
$\kappa_v \to \infty$ and $\lambda \to
 \pi k_F/\Lambda $ (constant), we find that
$L_2 \sim -ln x /x \to 0$. This gives $\dot{\gamma} \simeq (-b-\gamma)\gamma$ 
and $\gamma(t)$ ($\gamma(t)$ should be positive)
tends to vanishingly small. This contradicts the starting assumption.
Therefore, $\gamma(t)$ cannot tend to large in the IR limit.

(ii-3)  After all, the only possible situation is that 
$\gamma^*=0$, i.e., IR limit is the trivial fixed point. 
In this case, $\kappa_v$ and 
$\lambda$ become large (for $ 0 \leq b \leq 2 $) according to the original canonical scaling
such as
\beg
\dot{\kappa_v} &\simeq& \kappa_v,\no
\dot{\lambda} &\simeq& (2-b)\lambda.
\en

Since we have obtained the qualitative behavior  
of the parameters in the IR limit,
let us turn to the numerical calculations of the RG flows.
The result shows  not only that the above qualitative argument is correct
but also that some interesting crossover phenomenon appears.
We discuss the two cases (1) $0\le b <1$ and (2) $1 \le b \le 2$ separately.

(1) The case $0\le b <1$. 

When $\kappa_v=0$, as mentioned, $\gamma(t)$ has the nontrivial IR
fixed point $\gamma^*_0 \equiv 1-b > 0$, and at this fixed point the system 
behaves as  a non-Fermi liquid\cite{onoda,Nayak}.
On the line of $\kappa_v = 0$, the RG flows converges to this point. 
The RG flow whose initial value of $\kappa$ is very small,
 $\kappa_v \simeq 0$, first approaches the point $P_0 (\kappa_v  =0,
 \gamma = \gamma^*_0 > 0)$. Then  $\kappa_v$ increases and  the 
 flow  goes away from the point $P_0$,   finally  
approaching  the trivial fixed point  $ P_1 (\kappa_v  = \infty,
 \gamma = 0)$. 
Therefore, the point $P_0$ is an unstable
fixed point of this theory and the IR limit of the system is controlled 
by the point $P_1$ which describes a  Fermi liquid with impurity effects. 
Such behavior is shown in Fig.4. and Fig.5. 
In Fig.4 we plot the flows close to $\gamma(t)=1-b$. In Fig.5 we show 
the global behavior of  RG flows in  a wider region of $\kappa_v$.
In Fig.6 we show the wave-function renormalization constant $R(t)$,
together with the coupling constant $\gamma(t)$. 
As long as  $\gamma(t)$ is staying in the vicinity of  
the unstable fixed point $P_0$, 
$R(t)$ increases exponentially  $\sim \exp(2\eta t)$, where $\eta$
is the anormalous dimension of fermion field (for $\kappa_v  = 0$).
As discussed in Ref.\cite{onoda}, this behavior of $R(t)$
implies that the system in these energy scales is controlled by the 
unstable fixed point $P_0$ and it has non-Fermi-liquid-like behavior.
This result will be confirmed by the discussion on the fermion propagator
 which is given in Sec.6. 
As RG flow is going away from this unstable fixed point, approaching
the  trivial fixed point $P_1$, $\gamma$ decreases quickly and
 $R(t)$ becomes constant, as 
shown in Fig.6.

(2) $1 \le b \le 2$. 

The RG flows of this situation are shown in Fig.7. They reach the point
$P_1$ in the IR limit.
In both cases, $\kappa_v$ and $\lambda$ increase exponentially.

%
%
\section{Low-energy behavior of fermion propagator}
\setcounter{equation}{0}
%
%

In this section, let us consider the behavior of full fermion propagator
under the RG transformation with the generalized
scaling assignment (parameterized by $\xi$):
\begin{eqnarray}
\tilde{\omega}&=&e^{\xi t}\omega,\quad \tilde{p}=e^{t}p,\nonumber\\
\tilde{\psi}(a;\tilde{p},\tilde{\omega})
&=&e^{-(\xi + 1)t}\{R(t)\}^{\frac12}\psi(a;p,\omega).
\label{scalepsi1}
\end{eqnarray}
We begin with the definition,
\begin{eqnarray}
\langle\psi(a;p,\omega)\bar{\psi}(a';p',\omega') \rangle 
&\equiv&
-\delta_{a,a'}\bar{\delta}(p-p')\bar{\delta}(\omega-\omega')\nonumber\\
& &\hspace{2cm}
\times G( p, \omega; v_F,\gamma, \lambda, \kappa; \Lambda).
\label{scalepsi2}
\end{eqnarray}
By substituting the last line of Eq.(\ref{scalepsi1}) into the LHS of 
Eq.(\ref{scalepsi2}) and then expressing it by the propagator at $\tilde{p}$,
we obtain
\begin{eqnarray}
\langle\psi(a; p, \omega)\bar{\psi}(a'; p', \omega') \rangle 
&=& e^{2(\xi+1)t}\{R(t)\}^{-1}
\langle\tilde{\psi}(a; \tilde{p}, \tilde{\omega})
\bar{\tilde{\psi}}(a'; \tilde{p}', \tilde{\omega}') \rangle
 \nonumber\\
&=&-e^{2(\xi+1)t}\{R(t)\}^{-1}
\delta_{a,a'}\bar{\delta}(\tilde{p}-\tilde{p}')
\bar{\delta}(\tilde{\omega}-\tilde{\omega}')
\nonumber\\
& &\hspace{2cm} \times G(\tilde{p}, \tilde{\omega};
v_F(t), \gamma(t), \lambda(t),\kappa(t); \Lambda).
\label{scalepsi3}
\end{eqnarray}
From (\ref{scalepsi2}) and (\ref{scalepsi3}) we have 
\begin{equation}
G( p, \omega; \rho_i; \Lambda)
= e^{\xi t}\{R(t)\}^{-1}G(\tilde{p},
 \tilde{\omega}; \rho_i(t); \Lambda)
\label{scalepsi5}
\end{equation}
where $\rho_i(t)$'s denote the running parameters
$v_F(t),\gamma(t),\lambda(t)$ and $\kappa(t)$.
This equation can be used to evaluate the full fermion propagator simply by substituting
the solutions of the RG equations to the RHS of Eq.(\ref{scalepsi5}).

Explicitly, the   RHS of (\ref{scalepsi5}) is written as follows;  
\begin{eqnarray}
& &e^{-\xi t}R(t)G^{-1}(\tilde{p},
 \tilde{\omega}; \rho_i(t); \Lambda)\no
&\simeq & e^{-\xi t}R(t)
\Bigl[i\tilde{\omega} + i \kappa(t)  \sign(\tilde{\omega})
-   v_F(t)  e_{a}\cdot \tilde{p}  
-\Sigma_f(0, \tilde{\omega};\rho_i(t); \Lambda)
 \Bigl] \no
& \simeq&
 iR(t)\omega+iR(t)R_{\kappa}(t)\kappa \sign(\omega)
-R(t)R_{v_F}(t) v_F e_{a}\cdot p \no
& &-e^{-\xi t}R(t)
\Sigma_f(0, \tilde{\omega};\rho_i(t); \Lambda),
\end{eqnarray}
where $\Sigma_f(\tilde{p}=0,\tilde{\omega}; \rho_i(t); \Lambda)$ 
is the fermion self-energy which is obtained by integrating out {\em all the  modes} of fields (i.e., with momenta {\em smaller} than the cut off $\Lambda$). 
We have neglected its momentum-dependent part which is less dominant 
than $v_F e_{a}\cdot p$ at low energies. 
We evaluate the self-energy $\Sigma_f(0, \tilde{\omega};\rho_i; \Lambda)$
at one-loop level as follows;
\begin{eqnarray}
& &\Sigma_f(0, \tilde{\omega}; \rho_i(t); \Lambda)
\nonumber\\
& &=\frac{g^2(t)v_F^2(t)}{v_B\Lambda}
\int\underline{d\tilde{\epsilon}}\int^{\Lambda}\underline{d\tilde{q}}\;
 \left(\frac{e_a\times \tilde{q}}{\tilde{q}}\right)^2 \no
& & ~~~\times
 \frac{1}{i(\tilde{\omega}+\tilde{\epsilon})
+i \kappa \sign(\tilom+\tilep)
-v_F(t) e_{a}\cdot\tilde{q}}
 \cdot\frac{1}{\left(\frac{\tilde{q}}{\Lambda}\right)^{2-b}
 +\frac{\lambda(t)|\tilde{\epsilon}|}{v_F(t) \tilde{q}}}
 \nonumber\\
& &=-i\frac{g^2(t) v_F(t)}{(2\pi)^2 v_B \Lambda}
   \int d\tilde{\epsilon}\int^{\Lambda}_{0}d\tilde{q}\;
  \frac{\mbox{sgn}(\tilde{\omega}+\tilde{\epsilon})}
   {\left(\frac{\tilde{q}}{\Lambda}\right)^{2-b}
 +\frac{\lambda(t)|\tilde{\epsilon}|}{v_F(t) \tilde{q}}}\, \no
& &~~~\times 
\left[\sqrt{1+\left\{
\frac{|\tilde{\omega}+\tilde{\epsilon}
+ \kappa \sign(\tilom+\tilep)|}{v_F(t)\tilde{q}}
\right\}^2}
-\frac{|(\tilde{\omega}+\tilde{\epsilon})
+ \kappa \sign(\tilom+\tilep)|}{v_F(t)\tilde{q}}
\right],
\label{self}
\en
where  we have performed the angle integration, and used the fact
$\sign(\tilde{\omega}+\tilde{\epsilon}
+\kappa \sign(\tilom+\tilep))= \sign(\tilom +\tilep)$.
In the $\tilep$-integration above, the region of $\tilep \simeq 0$ gives 
dominant contribution, so  we treat  $\tilep$ as small 
and restrict the integration region   to $ 0 < \tilep < v_F(t)
\tilde{q}$.
We also treat  $\tilom$  small because we are  interested in
the low-energy behavior of the propagator.
Below, we discuss  the three cases, (I) $0<b<1$, (II) $ b = 1$, 
(III) $1 <b < 2$ separately.
 
{\bf (I) $0<b<1$}

In the previous section, we showed the RG flows of the parameters
 $\gamma(t),\kappa_v(t)$ and $\lambda(t)$.
In the small $\kappa_v(t)$ region, 
the expansion parameter $\gamma(t)$ approaches
 the saddle point $P_0$ with $ \gamma^*_0 = 1-b$. 
As $\kappa_v(t)$ tends to large,
the parameter $\gamma(t)$ leaves away from $P_0$  and
approaches the trivial fixed point $P_1$.
We study the low-energy  behavior  of the propagator in two regions;
(i)  region near the
saddle point $P_0$ where $\kappa_v(t)$ is small, and
(ii) region  near the trivial fixed points $P_1$ where $\kappa_v(t)$ is large. 

{\bf (i) Small $\kappa_v(t)$}

For small $\kappa_v(t)$,  the self energy is evaluated as
\beg
& &\Sigma_f(0, \tilde{\omega};\rho_i(t); \Lambda)
\no
& &= -i \frac{\gamma(t)}{2 \Lambda K_2(t)}
\int d\tilep \int_0^{\Lambda} d\tilde{q}
\frac{1}{\left( \frac{\tilde{q}}{\Lambda} \right)^{2-b}
+
\frac{\lambda(t) |\tilep|}{v_F(t) \tilde{q}}}
\sign(\tilom+ \tilep)
\left(
1-\frac{|\tilom +\tilep|+\kappa(t)}{v_F(t) \tilde{q}}\right)\no
& &~~~
+O\left(\tilom^2,\tilde{\kappa_v}^2(t)\right)\no
& &\simeq \Sigma_1 + \Sigma_2  + \Sigma_3
\en
where
\beg
\Sigma_1 &\equiv& 
-i \frac{\gamma(t)}{2 K_2(t)}
 \int d\tilep \; \sign(\tilom+\tilep)
\int_0^1 dq \frac{q}{q^{3-b} + 
\lambda_v(t) |\tilep| }\no
&=&
-i \frac{\gamma(t)}{ K_2(t)} \sign(\tilom) \int_0^{|\tilom|} d\tilep
\int_0^1 dq \frac{q}{q^{3-b} + 
\lambda_v(t) |\tilep|} \no
&=&
-i  \frac{\gamma(t)}{ K_2(t)} \tilom
(\lambda_v(t)|\tilom|)^{-\frac{1-b}{3-b}}
 H_{1b}(\lambda_v(t)|\tilom|) \no
\Sigma_2 
&=&
i  \frac{\gamma(t)}{ K_2(t)} \kappa_v(t) \tilom 
(\lambda_v(t) |\tilom|)^{-\frac{2-b}{3-b}}
H_{2b}(\lambda_v(t) |\tilom|) \no
\Sigma_3
&=&
i \frac{\gamma(t)}{ K_2(t)} \frac{\tilom}{v_F(t) \Lambda}
\int_0^{v_F(t) \Lambda} d\tilep
\int_0^1 dq \frac{1}{q^{3-b} + \lambda_v(t) \tilep}\no
&=&
i \frac{\gamma(t)}{ K_2(t)}
 \lambda(t)^{-\frac{2-b}{3-b}} \tilom
H_{2b}(\lambda(t) )
\label{selft} 
\en
and
\beg
\lambda_v(t) &\equiv& \frac{\lambda(t)}{v_F(t) \Lambda} \no
H_{1b}(c) &\equiv& \frac{1}{3-b}\int_c^{\infty}
dy y^{\frac{b-5}{3-b}} \ln (1+y) \no
H_{2b}(c) &\equiv& \frac{1}{3-b} \int_c^{\infty}
dy y^{\frac{b-4}{3-b}} \ln (1+y).
\en
The upper limit of integral region with respect of $\tilep$ in $\Sigma_3$ is
$v_F(t) \Lambda$ then the argument contained in   $\Sigma_3$ is $\lambda(t)$ and 
is not $\lambda_v(t)$.
To obtain an explicit form of the propagator, we need to evaluate
these $H$'s. 

Now we consider the case  where $t$ increases so that
 $\lambda(t)$ becomes large, while $\gamma(t)$ is still 
staying near the saddle point and 
 the argument $c=\lambda_v(t) |\tilom|$ of 
$\Sigma_1, \Sigma_2$ is still small.
Then,  we evaluate the above integrals for small $c$
\beg
H_{1b}(c) &\simeq&
\frac{1}{1-b}(1-c^{\frac{1-b}{3-b}}) \no
H_{2b}(c) &\simeq&
\frac{1}{2-b}(1-c^{\frac{2-b}{3-b}}).
\label{H12b}
\en
On the other hand,  the argument $\lambda(t)$ of $H_{2b}$ in $\Sigma_3$ is large, 
and then for large $c$, 
\beg
H_{2b}(c) \simeq  c^{\frac{1}{3-b}}(\ln c + 3-b).
\label{H12b2}
\en
We have used the assumption of small $\lambda_v(t) |\tilom|$ above just to evaluate
the integrals $H_{1b}$ and $H_{2b}$.
The discussion  later on shows that the Green function obtained below
by using (\ref{H12b}) is not only a solution to the RG equation at low energies
for small $\lambda_v(t) |\tilom|$ but also remains to be an approximate
solution even for large $\lambda_v(t) |\tilom|$.

By using (\ref{H12b}) and (\ref{H12b2}), we obtain the Green function as follows;
\beg
& &
e^{-\xi t}R(t)G^{-1}(\tilde{p},
 \tilde{\omega}; \rho_i(t); \Lambda)\no
& & \;\;
=i R(t) \omega + i e^{-\xi t}R(t)\kappa(t) \sign(\omega) -
 R(t)R_{v_F}(t)v_F  e_a \cdot  p  \no
& &~~~
- i e^{-\xi t} R(t) \frac{\gamma(t)}{K_2(t)}
\Biggl( \sign(\omega)
\left( \frac{\kappa_v(t)}{2-b} |\tilom|^{\frac{1}{3-b}}
\lambda_v(t)^{-\frac{2-b}{3-b} }
- \frac{1}{1-b} |\tilom|^{\frac{2}{3-b}}
\lambda_v(t)^{-\frac{1-b}{3-b} } \right)\no
& & ~~~
+ \tilom \left(\frac{\kappa_v(t)}{2-b}
-\frac{1}{1-b}  + \frac{1}{\lambda(t)}
(\ln \lambda(t)+3-b)\right) \Biggl).
\label{green}
\en
In the situation  where
 $\lambda(t)$ becomes large, while $\gamma(t)$ is still 
staying near the saddle point and $\kappa_v$ is small value,
 we obtain the folloing equation from (\ref{K1}) such as
\beg
K_1 &\simeq& K_2 =1-\frac{2 \kappa_v}{\pi}+ O(\kappa_v^2).
\en
From this relation and 
$R(t)$ , $R_{v_F}(t)$ in Eq.(\ref{rec}), we obtain 
\beg
R(t) &\simeq&  e^{\int_0^t ds \gamma(s)}\no
R(t)R_{v_F}(t) &\simeq& 1.
\en
The formal solution of $\gamma(t)$ which is given by (\ref{formalsol}) 
and the above equation give the following relation, 
\beg
R(t)\gamma(t) = \alpha_0 e^{(1-b)t} K_2(t),
\label{Ra}
\en
where $\alpha_0$ is the initial value of
 $\alpha(t)$ of (\ref{alpha}).
From (\ref{rge}), we also find 
\beg
v_F(t) &\simeq&e^{(\xi-1)t}v_{F0}\no
\kappa_v(t) &\simeq& e^t \kappa_{v0} \ll 1\no
\lambda_v(t) &\simeq& \lambda_{v0} e^{(3-b-\xi)t}.
\label{kalam}
\en
By using these equations, $R(t)$ is obtained as follows;
\beg
R(t) &=& 1+\alpha_0 \int_0^t ds e^{(1-b)s}K_2(s)\no
&=&1+\frac{\alpha_0}{1-b}
\left(e^{(1-b)t}-1\right)
-\frac{2\alpha_0 \kappa_{v0}}{\pi (2-b)}
\left(e^{(2-b)t} -1\right).
\label{Rt}
\en
We substitute these relations, which depend on $t$,
to Eq.(\ref{green}), and obtain the Green function as 
\beg
& &G^{-1}(p, \omega;\rho_i; \Lambda)
=e^{-\xi t}R(t)G^{-1}(\tilde{p},
 \tilde{\omega}; \rho_i(t); \Lambda)\no
& &  \simeq 
i\alpha \ \sign(\omega)\left(
\frac{\lambda_{v}^{-\frac{1-b}{3-b} }}{1-b}|\omega|^{\frac{2}{3-b}}
-\frac{\kappa_{v} \lambda_{v}^{-\frac{2-b}{3-b} }}{2-b}
|\omega|^{\frac{1}{3-b}} \right)\no
& &~~~+i\omega\left(1-\frac{\alpha}{1-b} +
\frac{2 \alpha \kappa_v}{\pi(2-b)}\right) + i\kappa \sign(\omega) 
-v_{F} e_a \cdot p +\Delta{G}^{-1}    \nonumber   \\
& & =\Big(G_{\mbox{\small leading}}\Big)^{-1}+O(\omega),
\label{G1}
\en 
where
\beg
\Delta{G}^{-1}=i\omega\alpha\left(e^{(2-b)t}\frac{2  \kappa_v}{\pi(2-b)}
(\frac{\pi}{2}-1)-\frac{e^{-bt}}{\lambda}
(\ln\lambda+t+3-b)\right),
\en 
and we have omitted the suffix $0$ for initial values of $\alpha, \lambda,\lambda_v,
\kappa_v$ and $v_F$.

Here we  comment on the last term $\Delta{G}^{-1}$ which depends on the RG
scale $t$.
The LHS of Eq.(\ref{G1}) is independent of the scale parameter  $t$,  
so the representation on the RHS should not have terms depending
on $t$ if we calculated the fermion self-energy exactly.
However one cannot do the exact calculation, and  we  evaluated
the self-energy perturbatively at one loop level assuming
 the RG parameters lie in the region where  $\gamma(t)\ll 1$.
If this region corresponds to the infrared limit (like the case
$\kappa_v \gg1$ and $1\leq b<2$ which we will discuss later),
the $t$-dependent terms should disappear in the infrared limit
($t \to \infty)$ with  exponentially dumping factors, for example see 
Eq.(\ref{G2}) and Eq.(\ref{G3}). 
In that case, the one-loop calculation is almost exact.
Right now, we are considering the situation with some finite $t$, 
$t \sim t^*$, 
in which  $\gamma(t)$ is staying 
near the saddle point $\gamma^*_0 = 1-b$. We assume $1-b \ll 1$ so that
the perturbative calculation assuming small $\gamma$ is meaningful.
However, the above $\Delta{G}^{-1}$ does not disappear even in the limit $t \to t^*$
with the $t$ dependent factor such as $e^{(2-b)t}$.
Of course, we find that the leading terms of Eq.(\ref{G1}), $G_{\mbox{\small leading}}$, 
are independent of  $t$ and 
are in fact a solution to Eq.(\ref{scalepsi5})
up to the first order of $\kappa_v(t)$ and at  low energies.
This fact suggests that the $\Delta{G}^{-1}$ term shall be canceled out
with  higher-loop corrections to the fermion self-energy.
Actually, the two-loop corrections are known to  contain  
terms of the order $(\gamma^*_0)^2/(1-b) = \gamma^*_0 $, 
which are of the  same order in $\gamma^*_0 = 1-b \; (\ll 1)$ as the one-loop correction. 

From (\ref{G1}), we find that the dominant terms of the Green function 
$\Big(G_{\mbox{\small leading}}\Big)^{-1}$
are $O(\omega^{\frac{2}{3-b}})$ and $O(\kappa_v\omega^{\frac{1}{3-b}})$ 
in the present  parameter region $0 < b <1$.
Therefore, at intermediate energy scales,
the Green function has the behavior of a non-Fermi liquid; a branch cut
rather than a pole in $\omega$ appears as in the Luttinger liquid in one dimension\cite{onoda}.

{\bf (ii) Large $\kappa_v(t)$}

In this case, the self-energy of (\ref{self}) is evaluated as 
\beg
& &\Sigma_f(0,\tilde{\omega};\rho_i(t);\Lambda)
\no
& &\simeq -i \frac{\gamma(t)}{2 \Lambda K_2(t)}
\sign(\tilom+\tilep)
\int d\tilep \int_0^{\Lambda} d\tilde{q}
\frac{1}{\left( \frac{\tilde{q}}{\Lambda} \right)^{2-b}
+
\frac{\lambda(t) |\tilep|}{v_F(t) \tilde{q}}}
\frac{v_F(t) \tilde{q}}{2\kappa(t)} \no
& &
=  - i\frac{\gamma(t)}{2\kappa_v(t) \lambda_v(t) K_2(t)}\sign(\tilom)
(\lambda_v(t)|\tilom|)^{\frac{3}{3-b}}
H_{3b}(\lambda_v(t)|\tilom|),
\en
where
\beg
H_{3b}(c) \equiv \frac{1}{3-b}
\int_c^{\infty} dy y^{\frac{b-6}{3-b}}
\ln(1+y).
\en
For small $c$, $H_{3b}$ is evaluated as 
\beg
H_{3b}(c) \simeq -\frac{1}{b}\left(
1-c^{\frac{-b}{3-b}} \right).
\en
Then  we obtain $G^{-1}(p, \omega;\rho_i; \Lambda)$ as 
\beg
& &e^{-\xi t}R(t)G^{-1}( \tilde{p},
 \tilde{\omega}; \rho_i(t); \Lambda)\no
& & \simeq
i R(t) \omega + i \kappa \sign(\omega) - v_F e_a \cdot p
+ie^{-\xi t}\frac{\pi R(t) \gamma(t)}{4 b }
\tilom \left( (\lambda_v(t) |\tilom|)^{\frac{b}{3-b}}
-1\right),
\en
where we have used $K_2(t) \simeq  2/(\pi \kappa_v(t))$.
Now $\kappa$ is large enough, so $R(t)$ behaves as
\beg
R(t) &\simeq& 1+\frac{2\alpha_0}{\pi}
\int_0^t ds \frac{1}{\kappa_v(s)}\no
&\simeq&
1+\gamma_0(1-e^{-t}),
\label{R(t)}
\en
where  $\gamma_0 \equiv 2 \alpha_0/(\pi \kappa_{v0})$.
Thus, in this case, the Green function is obtained as
\beg
& &e^{-\xi t}R(t)
G^{-1}(\tilde{p},\tilde{\omega}; \rho_i(t); \Lambda)\no
& &\simeq  i \kappa \sign(\omega) - v_F e_a \cdot p
+i\left(1+\gamma \right) \omega 
 -\frac{i \pi \gamma}{4b}
\lambda_v^{\frac{b}{3-b}}\sign(\omega)|\omega|^{\frac{3}{3-b}}
+ O(e^{-t})
\label{G2}
\en
where we have omitted the suffix $0$ as before.

In the low-energy limit, we find that the dominant term is linear 
in $\omega$, so the Green function has the Fermi-liquid behavior.
This result of course agrees with the conclusion  in Sec.5
obtained based  on the RG flow. 

{\bf (II) $b=1$}

In this case, $\gamma(t)$ goes to the trivial fixed point $P_1$.

{\bf (i) Small $\kappa_v(t)$}

 For small $\kappa_v(t)$, we obtain the self energy by setting 
$b=1$ in Eq.(\ref{selft}). The functions $H_{11}$ and $H_{21}$ are
easily obtained as
\beg
H_{11}(c) &=& \frac{1}{2}\left( (\frac{1}{c}+1)\ln(1+c)-\ln c \right) \no
H_{21}(c) &=& \frac{1}{\sqrt{c}}\ln(1+c) + 2 \arctan\frac{1}{\sqrt{c}}.
\en
From these equations and the $t$-dependences of parameters in (\ref{Ra}),
(\ref{kalam}) and (\ref{Rt}) with $b=1$, we obtain the Green function as 
follows;
\beg
& &e^{-\xi t}R(t)
G^{-1}(\tilde{p},\tilde{\omega}; \rho_i(t); \Lambda)\no
& &\simeq  i \kappa \sign(\omega) - v_F e_a \cdot p
+i(1+\frac{2\alpha \kappa_v}{\pi}) \omega 
+\frac{i \alpha}{2}\omega\ln \Big(\lambda_v|\omega| \Big)
+O(t)
\en
where   the suffix $0$ for initial values are omitted as before.
The last term $O(t)$ is the $t$-dependent part which should be canceled out
with higher-order corrections as explained above.

This expression contains $\omega ln(\omega)$ term, leading to
the weight of quasiparticles that vanishes logarithmically
 at low energies. This  behavior   
is like that of the marginal-Fermi liquid\cite{onoda}.

{\bf (ii) Large $\kappa_v(t)$}

 As $t$ becomes large,  $\kappa_v(t)$  becomes large too.
 Then we evaluate the self energy as
\beg
\Sigma(0,\tilde{\omega};\rho_i(t);\Lambda)
\simeq -i \sign(\tilom) 
\frac{\gamma(t)}{2 \kappa_v(t) \lambda_v(t) K_2(t)}
(\lambda_v(t) |\tilom|)^\frac{3}{2}
H_{31}(\lambda_v(t) |\tilom|).
\en
Here $H_{31}$ is evaluated as follows
\beg
c^{\frac{3}{2}}H_{31}(c) =
\frac{1}{3}\left(
\ln (1+c) + 2c(1 -\sqrt{c}\arctan \frac{1}{\sqrt{c}})
\right).
\en
By using (\ref{kalam}) and (\ref{R(t)}) with
$b=1$, we obtain the Green function as
\beg
e^{-\xi t}R(t)
G^{-1}(\tilde{p},\tilde{\omega}; \rho_i(t); \Lambda)
\simeq 
 i \kappa \sign(\omega) - v_F e_a \cdot p
+ i ( 1 + \gamma ) \omega +O(e^{-t}). 
\label{G3}
\en
The logarithmic term has disappeared but only the term linear in $\omega$
survives (together with the signature term). This  
behavior of the Green function is that of the ordinary Fermi liquid.

{\bf (III) $1<b < 2$}

In this parameter region, the Green function has the same form with (\ref{G1})
for $\kappa_v(t) \ll 1$ and (\ref{G2}) for $\kappa_v(t) \gg 1$.
But the parameter $b >1$ here, and so the dominant term of the Green function
is linear in $\omega$ at low energies.
Therefore, we find that the Green function has the Fermi-liquid behavior
in both cases of  $\kappa_v(t) \ll 1$ and  $\kappa_v(t) \gg 1$.

%
%
\section{ Concluding Remarks } 
%
%
\setcounter{equation}{0}

In this paper, we have given detailed studies on low-energy behavior of
 the system of nonrelativistic
fermions interacting with a dissipative gauge field by using the WRG.
Especially, we investigated the stability problem of the nontrivial IR fixed point in the clean system against impurities.
The $\kappa$-term in the fermion propagator, which is generated by the 
interaction between fermions and impurities, is introduced for this purpose.
We showed that this term makes the non-Fermi-liquid fixed point unstable,
and in the IR limit the effective gauge-coupling constant tends to vanish.
However in intermediate energy scales,  the system is controlled
by the unstable fixed point as a saddle point, exhibitting non-Fermi-liquid behavior for some parameter regions of $b$ and $\kappa$.
This result is confirmed by obtaining the fermion propagator by solving the
 RG equation and also by calculating the wave function renormalization constant.

It is also quite interesting to investigate finite-temperature effects in the present gauge-fermion system by the RG analysis; in particlular, the same stability problem.
This subject is under study and we plan to publish the result
as a paper subsequent to the present one\cite{takano}.

\vspace{1in}

{\large{\bf Acknowledgments}}

One of us (H.T.) is thankful for a kind hospitality of the institute of 
physics at University of Tokyo, Komaba.
The other (M.O.) would like to thank Japan Society for the Promotion of Science (JSPS) for partial financial support.
The numerical calculation was done in the computer center
of Joetsu university of education.

\eject
%
%
\appendix
\renewcommand{\theequation}{A.\arabic{equation}}
%
%
\setcounter{equation}{0}
\section{Taylor series of the self energy}\label{coeff}

In this Appendix, we make the Taylor expansion of the fermion self energy
(\label{selfenergy}) to obtain the result (\ref{ecoeff}).
It is assumed that $z_0 = \omega +E_p$ takes a  small value and
the self energy is expanded around $y=E_q$.
After the $y$ integration, only even functions of $y$ in $I_i$ 
remain, and then we neglect odd functions of $y$ in the expansion of $I_i$.
\beg
I_1(\omega,E,\kappa) &=& -\frac{1}{z_2+1}
\Bigl(\frac{1}{2}\ln\frac{(\omega+1)^2}{\kappa^2+E^2}
+ i\sign(E)(\frac{\pi}{2}-\arctan\frac{\kappa}{|E|}) \Bigl)\no
&=&c_0(\ky)+c_1(\ky)z_0+c_2(\ky)iE_p,\no
I_2(\omega,E,\kappa) &=& \frac{1}{z_1+1}
\Bigl(\frac{1}{2}(\ln\frac{(\omega+\kappa)^2+E^2}{\kappa^2+E^2}
+\ln|\omega+1|^2)\no
&-& i\sign(E)(\arctan\frac{\omega+\kappa}{|E|}-
\arctan\frac{\kappa}{|E|}) \Bigl)\no
&=& d_1(\ky)z_0+d_2(\ky)iE_p,\no
I_3(\omega,E,\kappa) &=& \frac{1}{z_1-1}
\Bigl(\frac{1}{2}(\ln((\omega+\kappa)^2+E^2)
+ i\sign(E)(\frac{\pi}{2}-\arctan\frac{\omega+\kappa}{|E|}) 
\Bigl)\no
&=&g_0(\ky)+g_1(\ky)z_0+g_2(\ky)iE_p.
\en
where
\beg
c_0(\ky)&=&\frac{1}{2a_-}\Bigl((1-\kappa)\ln(\kappa^2+y^2)
-2|y|(\frac{\pi}{2}-\arctan\frac{\kappa}{|y|}) \Bigl),\no
c_1(\ky)&=&\frac{1}{a_-}\Bigl( \kappa-1+D(\ky) \Bigl),\no
c_2(\ky)&=&-\frac{1}{a_-}\Bigl(
 (1-\kappa)(\frac{\kappa}{\kappa^2+y^2}-1)
+\frac{y^2}{\kappa^2+y^2} \Bigl)+E(\ky),\no
d_1(\ky)&=&-d_2(\ky)=\frac{\kappa}{\kappa^2+y^2},\no
g_0(\ky)&=&-c_0(\ky),\no
g_1(\ky)&=&\frac{1}{a_-}\Bigl(
\frac{(\kappa-1)\kappa-y^2}{\kappa^2+y^2}+D(\ky) \Bigl),\no
g_2(\ky)&=&E(\ky),
\en
and
\beg
D(\ky)&\equiv&\frac{1}{2a_-(\ky)}
\Bigl( -c_-(\ky)-\ln(\kappa^2+y^2) + 4(1-\kappa)|y|
(\frac{\pi}{2}-\arctan\frac{\kappa}{|y|}) \Bigl),\no
E(\ky)&\equiv&-\frac{2(1-\kappa)}{a_-(\ky)}\Bigl(
\frac{\pi}{2}-\arctan\frac{\kappa}{|y|} \Bigl),\no
a_-(\ky)&\equiv&(1-\kappa)^2+y^2,\no
c_-(\ky)&\equiv&(1-\kappa)^2-y^2.
\en

The self energy is expanden as follows;
\beg
\Sigma(a;p,\omega) =\Sigma_0+
\Sigma_{\omega} i\omega + \Sigma_p E_p.
\en
where
\beg
\Sigma_0 &\equiv&
 -i \alpha dt \FL \int^{\lambda}_{-\lambda} dy |\sin \phi(y)|
\sign(\omega)(c_0(\oky)+g_0(\oky))\no
&=& 0,\no
\Sigma_{\omega}
&\equiv&
-\alpha dt \FL \int^{\lambda}_{-\lambda} dy |\sin \phi(y)|
\Bigl(c_1(\oky)+d_1(\oky)+g_1(\oky) \Bigl)\no
&=&
\frac{1}{a_-(\oky)} \Bigl(
2(\olk-1)-\frac{c_-(\oky)}{a_-(\oky)}\ln(\oM)
+ \frac{4|y|(1-\olk)}{a_-(\oky)}
(\frac{\pi}{2}-\arctan\frac{\olk}{|y|}) 
\Bigl),\no
\Sigma_p &=& I_p-\Sigma_{\omega}.
\en
and
\beg
I_p&\equiv&
\alpha dt \FL \int^{\lambda}_{-\lambda} dy |\sin \phi(y)|
\Bigl(c_2(\oky)+d_2(\oky)+g_2(\oky) \Bigl)\no
&=&
\alpha dt \FL \int^{\lambda}_{-\lambda} dy |\sin \phi(y)|
\frac{-2\olk}{\oM},
\en
$y$ and $\olk$ are defined in Eq.(\ref{f1}) such as 
$y= v_F \Lambda \cos\phi / \FL $ and 
$\olk = \kappa / \FL $.
%
%
\section{Vacuum porlarization}\label{vacp}
\renewcommand{\theequation}{B.\arabic{equation}}
\setcounter{equation}{0}
%
%
In this Appendix, the vacuum porlarization (\ref{selfenergy2}) is calculated
to give (\ref{vacpol}).
In the $y$ integration, only even functions of $y$ in $P_i$ 
survive, and  odd functions of $y$ are omitted.
Then $P_i$'s are obtained  as follows,
\beg
P_1(\epsilon,E_q,y,\kappa)
&=&
\Bigl( \frac{1}{2}\ln \frac{L}{L_e} -\frac{\kappa}{L}iE_q
\Bigl),\no
P_2(\epsilon,E_q,y,\kappa)
&=&
\Bigl( -\ln \frac{L}{L_e} -(\frac{\kappa}{L}
-\frac{\epsilon+\kappa}{L_e})iE_q
\Bigl),\no
P_3(\epsilon,E_q,y,\kappa)
&=&
\Bigl( \frac{1}{2}\ln \frac{L}{L_e} -\frac{\epsilon+\kappa}{L_e}iE_q
\Bigl),
\en
where
\beg
L \equiv \kappa^2 + y^2,  L_e \equiv (\epsilon+\kappa)^2 + y^2.
\en
We assume that the coefficient of the impurity term $\kappa$ is larger than 
the momentum and energy, $q$ and $\epsilon$, as $q/\kappa \ll 1$ and
$\epsilon/\kappa \ll 1$. In this situation, we find that 
$P_1$ and $P_3$ are order $O(1/\kappa)$ and $P_2$ is order
$O(1/\kappa^2)$, and then we can neglect the $P_2$ term.
We  also set $\ln(L/L_e) \sim 0$ because
$L \sim L_e$.
After all we obtain the approximate form of $P_i$ as
\beg
\sum_{i=1}^3 P_i(\epsilon,E_q,y,\kappa)
&\simeq&
\frac{E_q}{2\pi(i\epsilon-E_q)}
(\frac{\kappa}{L}+\frac{\epsilon+\kappa}{L_e}),\no
&\simeq&
\frac{E_q}{\pi(i\epsilon-E_q)}\frac{\kappa}{L}.
\en
The vacuum porlarization part is given as follows
\beg
\Pi(q,\epsilon) &=& \frac{g^2 v_F \Lambda dt}{2 \pi^2}
\sum_a \frac{K_a E_q}{2\pi(i\epsilon-E_q)}
\int_0^1dx \frac{4}{\sqrt{1-x^2}}
\frac{\kappa_v}{\kappa_v^2+x^2}\no
&\simeq&
v_B \Lambda \alpha dt \frac{k_F}{2\Lambda}
\int^{\pi}_{-\pi}\frac{d\phi}{2\pi}\frac{v_F q \cos \phi \sin^2 \phi}
{i\epsilon-v_Fq\cos\phi}\int_0^1dx \frac{4}{\sqrt{1-x^2}}
\frac{\kappa_v}{\kappa_v^2+x^2}\no
&\simeq&
const + v_B \Lambda  \alpha dt \frac{ k_F}{2\Lambda}
\frac{|\epsilon|}{v_F q}\int_0^1dx \frac{4}{\sqrt{1-x^2}}
\frac{\kappa_v}{\kappa_v^2+x^2}\no
&\simeq&
const + 
 v_B \Lambda  \alpha dt \frac{\pi k_F}{\Lambda} \frac{|\epsilon|}{v_F q}
\frac{2}{\pi}\arctan\frac{1}{\kappa_v},
\label{vacpol2}
\en
where
\beg
K_a &\equiv& \Bigl(\frac{e_a \times q}{q} \Bigl)^2,
\en
and use the relation $\sum_{e_a}(\frac{2 \Lambda}{k_F}) \simeq
\int^{\pi}_{-\pi} d\phi$. In the x integration, we set 
$\sqrt{1-x^2} \simeq 1$.

\pagebreak

{\Large{\bf Figure Captions}}\\
\noindent
Fig.1  \\
Momentum distribution of the fermion density.
The existence of the $\kappa$-term smears the step-function-like behavior
at $\kappa=0$.
The energy scale corresponding to the momentum cut off $\Lambda$ is also
shown.  \\

\noindent
Fig.2 \\
Partition of the two-dimensional  space of fermion momentum $\vec{K}$.
The big arc $C-C'$ shows a part of Fermi circle of radius $K_F$ around
the origin $O$. Theare are $N=2\pi K_F /(2\Lambda)$ Fermi wave vectors,
 $\vec{K}_{F,a} (a=1,..., N)$, being separated each other by   equal  
distances $2\Lambda$ along the circle. Here $\Lambda$ is the  cutoff momentum. 
Under the renormalization-group transformations,
we integrate gradually the modes with the momenta that sit ``near" the Fermi circle within the distance $\simeq \Lambda$. 
To be explicit, we integrate the momenta wich locate within
one of the $N$ small circles of radius $\Lambda$,
each circle being around the end point of $\vec{K}_{F,a}$ as illustrated. 
One may enlarge the  region of relevant momenta, e.g., to those that locate
within the union of small circles  of radius $2\Lambda$, to avoid
the momenta which sit very near to the Fermi circle but  never be integrated.
However, the main conclusions of the text do not depend on such detail.
\\

\noindent
Fig.3 \\
The two-dimensional  space of momentum $\vec{k}=(k_x,k_y)$.
$\vec{k} = \vec{K}-\vec{K}_{F,a}$ is a deviation of fermion momentum 
$\vec{K}$ near the Fermi circle measured from the nearest Fermi momentum $\vec{K}_{F,a}$.
In the step of renormalization-group transformation from $t $ to $t + dt$,
the modes with such momenta that sit in the shaded area  are  to be integrated.
This area is nothing but a shell surrounded by two circles of radii $\Lambda 
e^{-t}$ and $\Lambda e^{-(t+dt)}$.   \\

\noindent
Fig.4  \\
Structure of RG flows of $\kappa_v(t)$ and $\gamma(t)$ near the
unstable fixed point $P_0$ $(\gamma^{\ast}, \kappa_v)=(1-b,0)$. 
Values of the parameters are $b=0.5, \; k_F=1, \; \Lambda=1$ and $\lambda=0.1$.
For sufficiently small initial value of $\kappa_{v0}$, the 
RG flows approach to the fixed point $P_0$.
\\

\noindent
Fig.5  \\
Global structure of the RG flows. Same as Fig.4, but the scale of $\kappa_v$
is enlarged.
The effective coupling constant $\gamma(t)$ tends to vanish in the IR limit.
\\

\noindent
Fig.6  \\
The solid line is the effective coupling constant $\gamma(t)$, and the dashed line
is the renormalization constant $\ln R(t)$.
As long as  the effective coupling constant stays near the unstable fixed point
$P_0$,
the renormalization constant grows up exponentially.
\\

\noindent
Fig.7  \\
RG flows for $1\leq b \leq 2$.
$\kappa_v(t)$  increase exponentially and $\gamma(t)$ tends to vanish in the IR limit. 

\end{document}